\documentclass[aps,final,reprint,twocolumn,floatfix,notitlepage]{revtex4-1}
\usepackage{graphicx,amssymb,soul,color,amsmath,bm}
\usepackage{epstopdf,hyperref}
\usepackage{braket,dsfont}
\usepackage[mathcal]{euscript}

\hypersetup{colorlinks=true,citecolor=blue,linkcolor=magenta}

\sethlcolor{yellow}
\allowdisplaybreaks

\begin{document}

\title{Excitonic density-waves, bi-excitons and orbital selective pairing in two-orbital correlated chains}
\author{Chun Yang}
\affiliation{Department of Physics, Northeastern University, Boston, Massachusetts 02115, USA}
\author{Adrian E. Feiguin}
\affiliation{Department of Physics, Northeastern University, Boston, Massachusetts 02115, USA}

\date{\today}
\begin{abstract}
We present a comprehensive study of a one-dimensional two-orbital model at and below quarter-filling that realizes a number of unconventional phases. In particular, we find an excitonic density wave in which excitons quasi-condense with finite center of mass momentum and an order parameter that changes phase with wave-vector $Q$. In this phase, excitons behave as hard-core bosons without charge order. In addition, excitons can pair to form bi-excitons in a state that is close to a charge density-wave instability. When pairing dominates over the inter-orbital repulsion, we encounter a regime in which one orbital is metallic, while the other forms a spin gapped superconductor, a genuine orbital selective paired state. 
All these results are supported by both, analytical and numerical calculations. By assuming a quasi-classical approximation, we solve the three-body hole-electron-spinon problem and show that excitons are held together by forming a bound state with spinons. In order to preserve the antiferromagnetic background, excitons acquire a dispersion that has a minimum away from $k=0$. The full characterization of the different phases is obtained by means of extensive density matrix renormalization group calculations.
\end{abstract}

\maketitle

\section{Introduction}

Charge recombination and photo induced charge transfer lie at the heart of current attempts to construct viable optoelectronic devices using organic semiconducting devices \cite{tang1987,burroughes1990,guftafsson1992}. 
In particular, a great deal of interest has focused on one-dimensional(1D) materials due to the band edge singularities that could give rise to a high-differential optical gain.
One dimensional materials such as conjugated polymers\cite{polymersBook,polymers_review} have already found uses in a wide range of applications such as light-emitting diodes, lasers, sensors, and molecular switches \cite{brown1995,burroughes1990,dodabalapur1995,dodabalapur1995b,hide1996,yang1998,schmitz2001,nitzan2003}.

%Even though these technologies are dominated by semiconductor materials, the main requirement is to possess a gap in the optical range. It is possible, and even desireable, to explore the optical properties of correlated materials, where the gap arises as a result of electronic interactions.
%While much theoretical work has focused on single-band problems, a rich phenomenology can occur in more realistic multi-orbital cases, where besides Coulomb interactions, Hund physics plays an important role. Amongst other important correlation-driven phenomena one could cite orbital selective Mott transitions\cite{Koga2004,deMedici2009,Vojta2010}, spin-orbital separation\cite{Wohlfeld2008,Wohlfeld2011,Wohlfeld2013,Schlappa2012,Chen2015}, spin-incoherent behavior\cite{Feiguin2011,soltaniehha2012}, and pair density waves\cite{Zachar2001,Berg2010}.
%Light-induced phenomena in two-orbital models has been studied focusing mainly in quarter-filled Mott insulators using double-exchange models \cite{Kanamori2009,Maeshima2010,Kanamori2011,Kanamori2012}. 

Excitons in low-dimensional strongly correlated electronic materials have received much theoretical attention
\cite{jeckelmann2000,tsutsui2000,essler2001,jeckelmann2003,gallagher1997,barford2002,Gebhard1997}, and they have also been observed experimentally in 1D Mott insulators \cite{ono2005,Schlappa2012}.
The behavior of excitations in interacting 1D systems is very peculiar: due to the pervasive nesting at all electron densities, Fermi liquid picture breaks down giving rise to a different paradigm, the Luttinger liquid (LL). In a spin-full Luttinger liquid elementary degrees of freedom are not fermions with well defined charge and spin, but bosonic collective quasi-particles carrying spin (spinon) and charge (holon), leading to the concept of spin-charge separation. 

Excitonic instabilities in multi-orbital systems typically arise as photo induced excitations and give rise to a complex interplay between charge, spin and orbital degrees of freedom \cite{Schlappa2012,Bisogni2013}. Understanding this interplay and how these bosonic excitations decay is one of the main goals of pump-probe spectroscopy. 
While much theoretical work has focused on single-band problems, a rich phenomenology can occur in more realistic multi-orbital cases, where besides Coulomb interactions, Hund physics plays an important role. Amongst other important correlation-driven phenomena one could cite orbital selective Mott transitions\cite{Koga2004,Rice2005,deMedici2009,Vojta2010}, spin-orbital separation\cite{Wohlfeld2008,Wohlfeld2011,Wohlfeld2013,Schlappa2012,Chen2015}, spin-incoherent behavior\cite{Feiguin2011,soltaniehha2012}, and pair density waves\cite{Zachar2001,Berg2010,Fradkin2015}.

%Light-induced phenomena in two-orbital models has been studied focusing mainly in quarter-filled Mott insulators using double-exchange models \cite{Kanamori2009,Maeshima2010,Kanamori2011,Kanamori2012}. 
%While much theoretical work has focused on single-band problems, a rich phenomenology can occur in more realistic multi-orbital cases, where besides Coulomb interactions, Hund physics plays an important role. Amongst other important correlation-driven phenomena one could cite orbital selective Mott transitions\cite{}, spin-orbital separation\cite{}, spin-incoherent behavior\cite{Feiguin2011,soltaniehha2012}, and pair density waves\cite{Zachar,Berg}.

Wannier-Mott excitons in semiconductors and their subsequent condensation are well understood since the 60's\cite{exciton_book1,exciton_book2,Hanamura1977,Halperin1968}. In strongly correlated systems one finds Frenkel, Mott-Hubbard excitons, and the recently proposed Hund excitons\cite{Rincon2017a,Nunez2017}, which are more tightly bound objects. Excitonic condensation in strongly correlated models has been studied in a number of scenarios \cite{Kunes2015}.
Early in this area of research it was pointed out that in multi-band Mott insulators not only the spin and charge can order, but also the orbital degree of freedom\cite{Kugel1982}. In one dimension one encounters that the associated excitations (orbitons) may also decouple from the spin in what is referred-to as ``spin-orbital'' separation. One way to understand this phenomenon is by starting from the simplest model describing a Mott insulator and accounting for both spin and orbital degrees of freedom, the Kugel-Komskhii (KK) chain \cite{Kugel1982}. It has been shown that the problem of a propagating orbiton can be mapped onto the dynamics of a hole in an antiferromagnet\cite{Wohlfeld2008,Wohlfeld2011,Wohlfeld2013,Schlappa2012,Chen2015}. This leads to an effective $t-J$ model which is much simpler and has extensively been studied in the literature. In one-dimension, the physics is described in terms of LL theory, which naturally explains spin-orbital separation.

In this work we study a more general problem that, in addition to orbital and spin degrees of freedom, also accounts of charge fluctuations. Our model bares resemblance to the so-called two-orbital Hubbard model, also referred-to as electron-hole Hubbard model. 
This problem has been extensively studied in higher dimensions, and also in 1D \cite{Nonne2010,Kaneko2012a,Okumura2001,Ogawa2005a,Ogawa2007a,Ogawa2008b,Asano2008b,Ogawa2009,Asano2013,Li2016,Patel2017}. Here we consider a modified version of it that applies in the strong coupling limit and we derive, both theoretical and numerically, a number of important results that highlight the non-trivial nature of the excitations. We analize the case of bound electron-hole pairs and spinons, and the eventual deconfinement of the excitations in one-dimension. We show that the non-trivial dispersion of the spinons leads to the formation of an excitonic condensate with finite center of mass momentum, and bi-excitons. 
The coexistence of a excitonic density waves with a ``normal'' electronic sea resembles the case of a Fulde-Ferell-Larkin-Ovchinnikov (FFLO) superconductor\cite{Fulde1964,larkin1964}, with unpaired electrons concentrating at the nodes of the oscillating condensate. %: a 1D quasi-supersolid. 
%In a supersolid, the pinning of the atomic density forms a regular
%structure (a crystal) accompanied by a superfluid of atoms (bosons)  of the same
%species (for a review see Ref.~\onlinecite{Boninsegni2012}), spontaneously breaking translational symmetry. 
%In a similar fashion, a condensate of Cooper pairs in a magnetic field (or an imbalanced Fermi mixture) can condense with a finite center of mass momentum with the excess polarized fermions sitting at the nodes of the superfluid order parameter. 
Unlike the conventional FFLO state, our model supports an excitonic condensate with a finite momentum $Q$, while the normal electrons behave as a fluid, without breaking translational symmetry.

The paper is organized as follows: In section \ref{intro} we introduce the model and describe certain limits; in section \ref{electron-hole-spinon} we solve the three-body problem of an electron-hole pair and a spinon, offering a rigorous and intuitive picture for the formation of bound states with finite center of mass momentum; in section \ref{results} we provide numerical support to our analysis using the density matrix renormalization group method (DMRG)\cite{White1992,White1993,schollwock2005density,DMRGbook,Feiguin2013a}. We conclude with a discussion of the results.

\section{Two-orbital $t-J$ model}\label{intro}

We consider a two-orbital $t-J$ model described by the Hamiltonian:
\begin{eqnarray}
H &=& -t\sum_{i,\sigma,\lambda}\left(c^\dag_{i\sigma\lambda}c_{i+1\sigma\lambda}+h.c.\right)
+ U'\sum_i n_{i1}n_{i2} \nonumber \\ 
&+& J \sum_{i,\lambda} \left( \vec{S}_{i,\lambda} \cdot \vec{S}_{i+1,\lambda} - \frac{1}{4}n_{i\lambda}n_{i+1,\lambda} \right) \\
&+& \Delta \sum_i \left(n_{i2}-n_{i1}\right) \nonumber
\end{eqnarray}
where $c_{i,\lambda,\sigma}^{\dagger}$ is a fermionic creation operator acting on a site $i$ and orbital $\lambda$ ($\lambda=1,2$) with spin $\sigma=\uparrow,\downarrow$, and the constraint forbidding double occupancy is implicit as usual. The operators $n_{i,\lambda}$ represent the local density while $\vec{S}_{i,\lambda}$ refer to the local spin.  The hoppings along the two legs $t$ are taken to be equal for simplicity, and to be our unit of energy, implying that for large $\Delta$ the model will display an indirect gap. In addition, we include a Coulomb repulsion between electrons on both orbitals parametrized by $U'$, and a Heisenberg interaction between fermions on the same orbital chain. We have ignored the Hund coupling and interchain hopping since, for instance, in Sr$_2$CuO$_3$ this is one order of magnitude smaller than the on-site Coulomb repulsion\cite{Neudert2000}. By analogy, this model represents strongly interacting electrons on two parallel chains interacting via an electrostatic Coulomb repulsion and is a well defined limit of the two-orbital Hubbard model at half-filling with $J=4t^2/(U+U')$. We consider the total number of electrons to be constant, and a crystal field splitting $\Delta$ determines the relative population of the two bands in the ground state. Clearly, the total spin $S^z$ and the number of electrons $N$ are conserved, but also $N_1$, $N_2$, $S^z_1$ and $S^z_2$ on each orbital chain are conserved independently. This means that $N_2= N-N_1$, and the last term of the Hamiltonian becomes just a constant shift:
\[
\Delta \sum_i \left(n_{i1}-n_{i2}\right) = \Delta(2N_1-N),
\]
which tells us that the crystal field splitting acts basically as a chemical potential for orbital excitations. The number of particle-hole pairs in the ground state could be arbitrarily tuned by changing $\Delta$, or by creating photo-induced excitations (notice that for this mechanism to be applicable, and inter-orbital hopping needs to be included). For $\Delta=0$ one obtains $N_1=N_2$, while for $\Delta > 2t$, $N_2 = 0$. Regardless, one could independently fix $N$ and $N_1$. Clearly, the case $N_1=N_2=L$ (half-filling) describes two independent Heisenberg spin chains without charge fluctuations. In the following we focus on the case $N <= L$, or density below quarter-filling.

We can gain some basic intuition on the problem by looking at three particular cases. First we consider $J=0$: In the absence of spin interactions, this degree of freedom becomes spurious. We can map each band onto a pseudospin quantum number, and identify $\lambda=1(2) \rightarrow \sigma=\downarrow(\uparrow)$. The problem is now equivalent to a one dimensional, single band Hubbard chain with $U' \rightarrow U$ and a magnetic field $2\Delta$. If we assume $\Delta=0$,
quarter filling corresponds now to half-filling and the ground state is an unpolarized Mott insulator. Creating an exciton by applying $c^\dagger_2 c_1$ can now be understood as $S^+=c^\dagger_\uparrow c_\downarrow$. The Mott insulating Hubbard chain has no spin gap, and therefore this costs no energy. However, the single particle spectrum is gapped in the charge sector. The charge gap can now be associated to the binding energy that holds the exciton together: it costs an energy of the order of $U'$ for an up particle to hop to a neighboring site already occupied by a down particle. However, away from quarter filling this is no longer the case and there are empty sites the up-particle can hop to. In this situation both spin and charge are gapless, the system becomes a Luttinger liquid, and particles and holes move freely. 

A second limit corresponds to $U'=0$: this maps onto two decoupled $t-J$ chains, and excitons are not stable quasi-particles. The ground state for this model has been extensively studied \cite{Moreno2011}. For large $J/t >2$ and intermediate densities the ground state presents dominant pair-pair correlations that decay algebraically. This indicates the formation of quasi-condensate (actual superconductivity is not realized in 1D and correlations decay algebraically), which has to be distinguished from an excitonic quasi-condensate. Therefore, by introducing a crystal field splitting, one band can realize pairing, while the other one remains a metal!

Finally, for finite $J$ and large $U'$ and $\Delta$ at quarter filling we find that a single exciton is strongly bound and the particle-hole pair can move coherently through high order processes. This particular scenario can be identified with the motion of a single hole in an antiferromagnet\cite{Wohlfeld2011}. However, the case at finite exciton density that occupies our attention in this study does not allow for such a simple interpretation and a deeper description of this regime is still lacking.

\begin{figure}%[ht]
\centering
\includegraphics[width=0.25\textwidth]{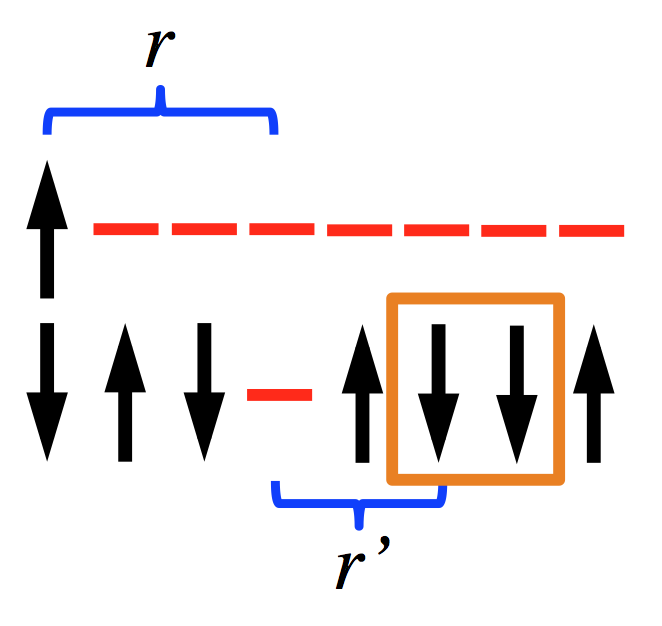}
\caption{Cartoon describing a typical state with the excited electron at the origin, a hole at a relative distance $r$, and a spinon at a relative distance $r'$ from the hole.}
\label{fig:exciton}
\end{figure}

\section{The electron-hole-spinon problem}\label{electron-hole-spinon}

\subsection{Single exciton}
To develop some intuition on the nature of the exciton condensate in this model we study a toy problem of one single exciton in the limit of strong uni-axial anisotropy (Ising). We assume that the system is at quarter-filling with one electron per site, and the band splitting $2\Delta$ is larger than the bandwidth $4t$. In this situation, a single band is half-filled, and the ground state is just an Ising antiferromagnet. We now create an exciton by promoting an electron to the upper band, and thus creating a hole in the lower band. It is intuitive to see that the Coulomb repulsion $U'$ acts as an attractive potential between the electron and the hole, and thus they will form a bound state. The formation of a two particle bound state in Hubbard-like models is a simple problem that has been studied in a number of setups in the literature \cite{Valiente2008,Valiente2009,Nguenang2009,Qin2008,Kato2012,Degli2014,Rausch2016,Rausch2017} (see an interesting analogy with phonons in Ref.~\onlinecite{Sous2017}). However, in our scenario, the situation is more complex, since the motion of the hole will leave behind a misaligned spin, a domain wall or spinon, that costs an energy $J$. Therefore, our analysis should also account for the presence of this defect in what now becomes a three-body problem in 1D. As complicated as it may sound, it turns out to be tractable, as follows. We consider a basis of states characterized by the position of the electron, $r_e$, the position of the hole relative to the electron $r = r_h-r_e$, and the position of the domain wall relative to the position of the hole $r'=r_s-r_h$, as illustrated in Fig.~\ref{fig:exciton}.

\begin{eqnarray}
H|r_e,r,r'\rangle &=& -t (|r_e+1,r-1,r'\rangle+|r_e-1,r+1,r'\rangle \nonumber \\
&+& |r_e,r+1,r'-1\rangle+|r_e,r-1,r'+1\rangle ) \nonumber \\
&+& U'\delta_{r,0}|r_e,r,r'\rangle \\ 
&-& J\delta_{r',-1}|r_e,r,r'\rangle \nonumber \\
&+& J(|r_e,r,r'+2\rangle+|r_e,r,r'-2\rangle). \nonumber
\end{eqnarray}
We assume periodic boundary conditions, which allows us to construct a basis of states that are translational invariant and labeled by a momemntum $k$:
\begin{equation}
|r,r',k\rangle = \frac{1}{\sqrt{L}}\sum_{x=0}^{L-1} e^{ikx} T_x |r_e=0,r,r'\rangle.
\end{equation}
Within each momentum sector we can easily obtain the Hamiltonian matrix elements
%, as shown in the Supplementary Material, 
and numerically diagonalize the problem for very large chains. 
In the $J=0$ limit, we should recover the results for two particles without spinon, and observe a band of bound states with a minimum at $k=0$ for sufficiently large $U'$. 
Our intuition tells us that if the binding energy is smaller than the kinetic energy of a free electron and a free hole, we will not obtain bound states. 

\begin{figure}%[ht]
\centering
\includegraphics[width=0.35\textwidth]{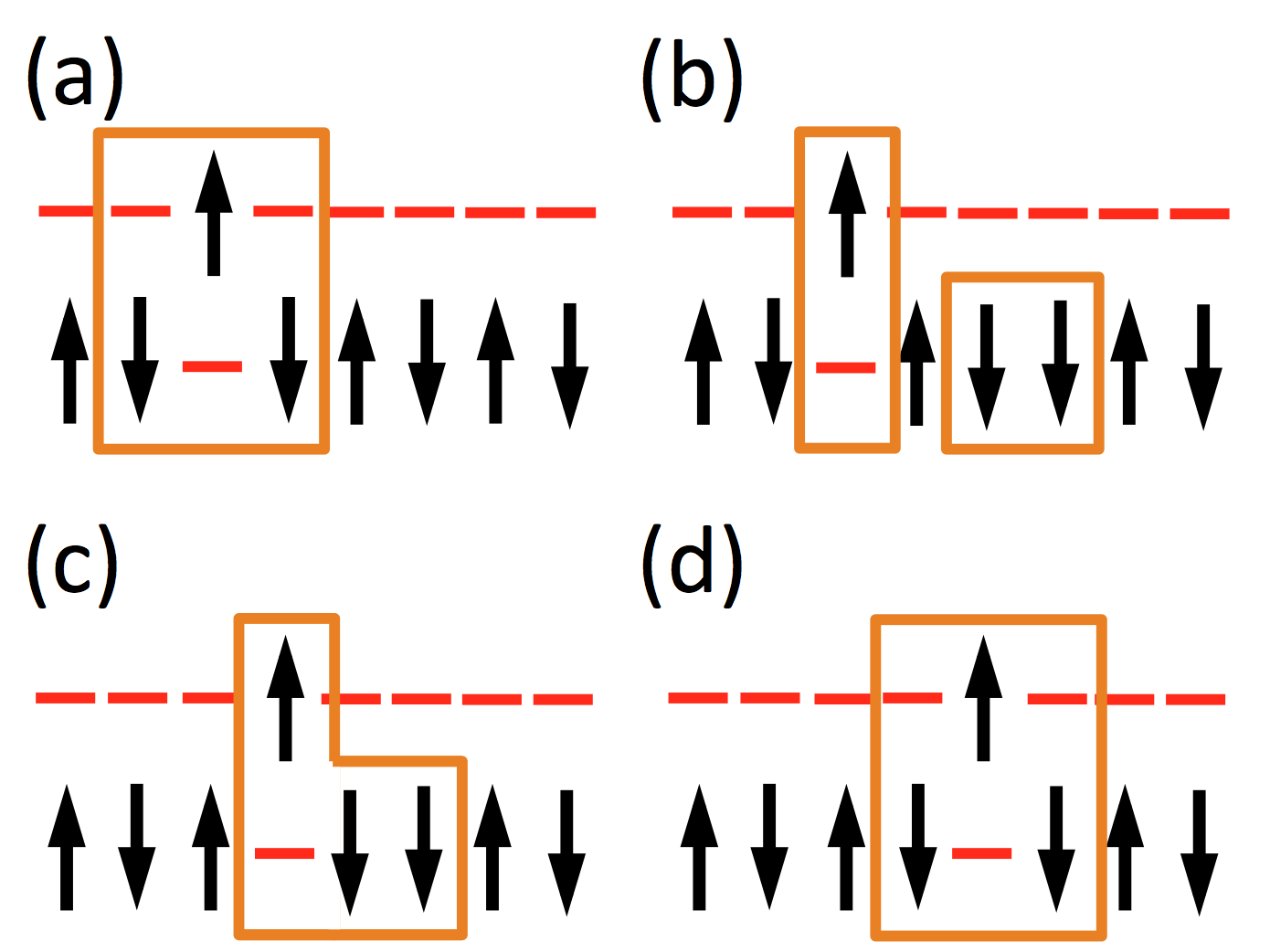}
\caption{Cartoon describing the high-order effective hopping of an exciton-spinon bound state. Each spin flip shifts the spinon by two lattice spaces. The exciton moves to remove the magnetic domain wall. We show the spinon moving as a single object.}
\label{fig:exciton_hopping}
\end{figure}

\begin{figure}%[ht]
\centering
\includegraphics[width=0.45\textwidth]{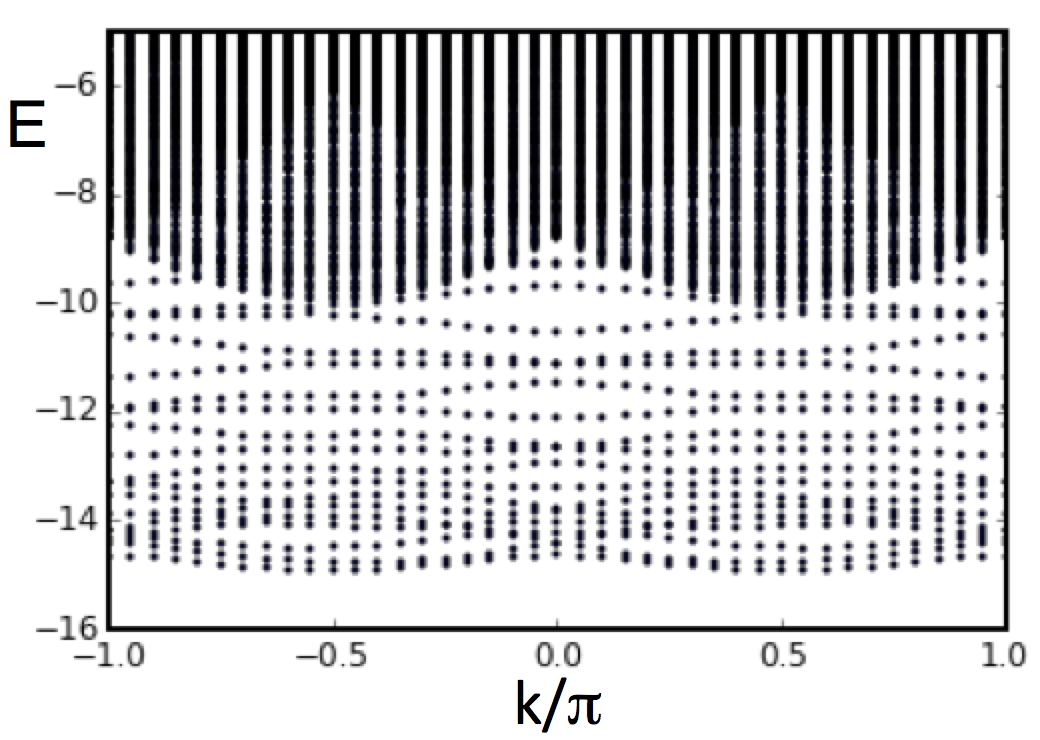}
\caption{Excitation energies for the electron-hole-spinon problem. The lowest energy band has a double-dip dispersion with minima at $k=\pm\pi/2$. The system size is $L=40$ and we used $U'=8$; $J=6$ to enhance the main features.}
\label{fig:energy_exciton}
\end{figure}

After introducing $J$, it is easy to see that the bound electron-hole pair behaves as a hole in the antiferromagnet that propagates coherently. This is the main idea behind the mapping to an effective $t-J$ model\cite{Wohlfeld2011}. For sufficiently large values, the free spinon and the electron-hole pair will also form a bound state, where the domain wall will be `absorbed' by the excitation. In order to account for the spin fluctuations we assume the approximation used in the seminal paper by Villain \cite{Villain1975} and we consider only spin-flip processes that move the domain wall, and ignore those that create new ones for being energetically too costly. It is easy to see that the spinon propagates by {\it two} sites for each spin-flip (see Fig.~\ref{fig:exciton_hopping}(b)), and therefore it has a dispersion $\epsilon_s(k)=2J\cos{(2k)}$. The larger mass of the spinon will tend to localize the electron-hole pair giving it a quite flat dispersion. However, it can still move without leaving a domain wall, as shown in Fig.~\ref{fig:exciton_hopping}. In order for this to happen, the bound state has to hop by two lattice spaces accompanied by a spin-flip, such that the resulting motion does not distort the antiferromagnetic background. This high-order processes allow for the spinon-electron-hole object to propagate with an effective second neighbor hopping, leading to a minimum in the dispersion at $k=\pm \pi/2$ and a maximum at $k=0$ (see Fig.\ref{fig:energy_exciton}). 

As a hint of what this means let us consider a finite density of excitons. These electron-hole pairs are now bosons that can condense with momentum $k=\pm \pi/2$. This condensate can break $Z_2$ symmetry by choosing one of the two momenta, or more likely form an equal superposition. This would correspond to a an order parameter that would oscillate in space as $\Delta_{cond} \sim \cos{(\pi x/2)}$.

We cannot forget that this picture assumes a classical magnetic ordering. In the isotropic $SU(2)$ limit the spinon forms a deconfined excitation and propagates independently, giving rise to the spin-orbital separation picture.

\begin{figure}%[ht]
\centering
\includegraphics[width=0.48\textwidth]{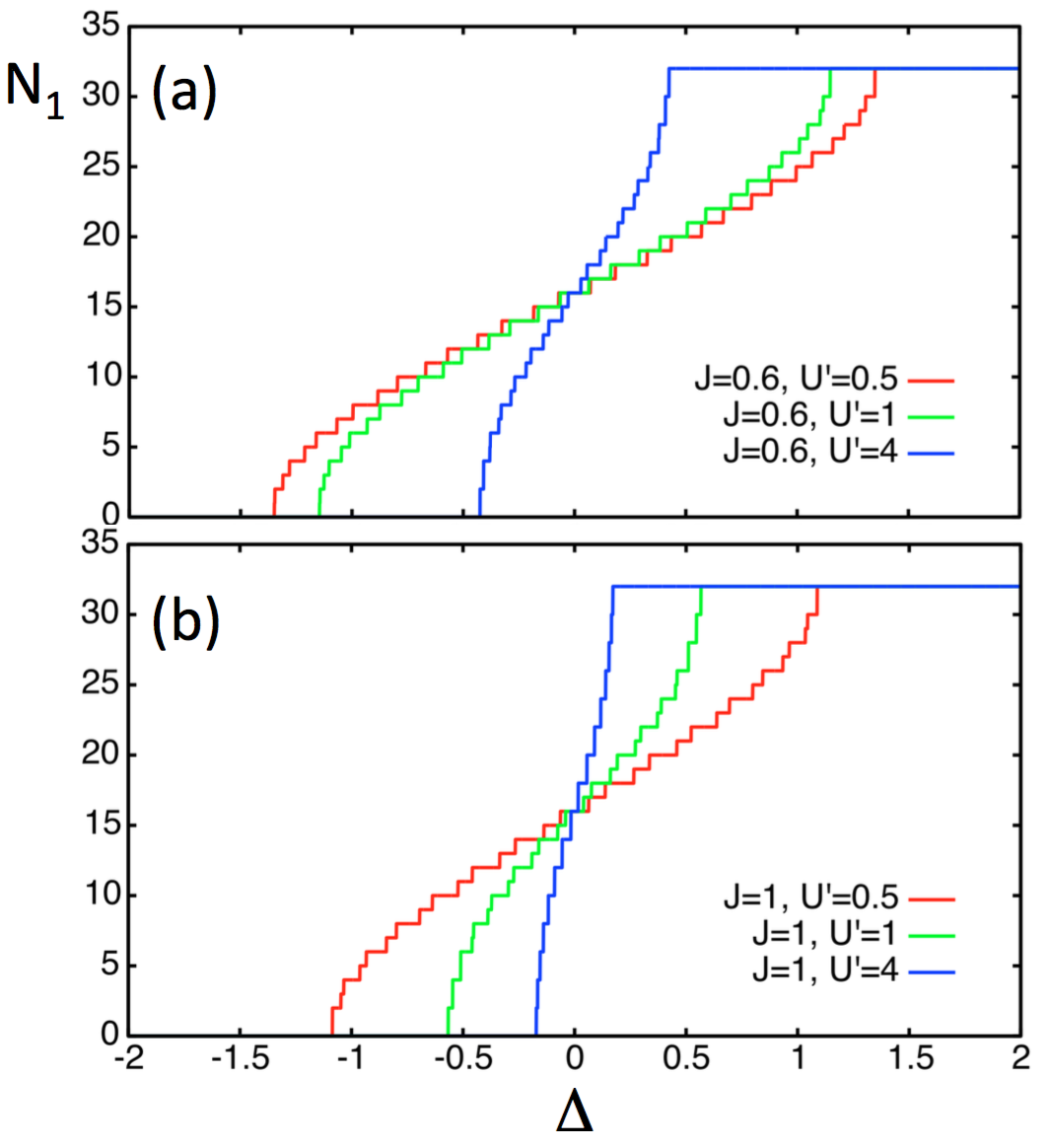}
\caption{Ground state occupation $N_1$ of the first orbital chain as a function of the band splitting $\Delta$ for a chain of length $L=32$. The total density is quarter filling and the occupation of the second chain is given by $N_2=L-N_1$. (a) Results for $J=0.6$ and (b) $J=1$ and several values of $U'$. The bi-exciton instability is signaled by jumps in steps of two.}
\label{fig:mu}
\end{figure}

\subsection{Bi-excitons and phase segregation}

A low density of excitons corresponds to a low density of electrons in the upper band. As it follows from the phase diagram of the 1D $t-J$ chain \cite{Moreno2011}, for sufficiently large values of $J$ and at low densities the ground state of the model becomes superconducting with a quasi-condensate of singlets held together by a binding energy of the order of $J$. In our model such pairs will be formed by excitons, {\it i.e.} they will form bi-excitons that are bound by an energy dictated by $J$ in the upper band, and $J$ in the lower band. Therefore, it is to expect that for moderate values of $J \sim t$ the system will realize a quasi-condensate of bi-excitons. Notice that this argument does not prevent the formation of excitonic `strings', where the excitons clump together forming a separate domain.
This would give rise to phase segregation, and would be manifested by a region of instability in the phase diagram where excitons and conduction electrons are spatially separated. This occurs when the interaction $U'$ is large and the excitons become very heavy. In this case it is easy to see that electrons on each band will form a string coupled only via the Heisenberg exchange term, and will occupy distinct regions of space, hence behaving as two independent Heisenberg chains.

%\begin{figure}%[ht]
%\centering
%\includegraphics[width=0.45\textwidth]{energy_biexciton.png}
%\caption{Excitation energies for the two exciton problem. The lowest energy band signals the presence of bi-excitons. The toy-model assumes two interacting hard-core bosons in $L=120$ sites with first and second neighbor hoppings $t_1=1; t_2=-0.5$, and attraction $U=-2$.}
%\label{fig:energy_biexciton}
%\end{figure}

\section{Numerical results}\label{results}
\subsection{Ground state}

We conduct density matrix renormalization group (DMRG) calculations for chains up to $L=64$ with open boundary conditions and keeping the truncation error below $10^{-6}$, which requires of the order of 2000 states in some cases. Most results, unless otherwise stated, correspond to $L=64, N_1=48$ and $N_2=16$, with Fermi momenta $k_{F1}=3\pi/8$ and $k_{F2}=\pi/8$, respectively. 
We first analize the $N_1$ vs. $\Delta$ for different values of the interaction $U'$ and $J$, as shown in Fig. \ref{fig:mu}. We ran the simulations in the canonical ensemble with fixed values of $N_1$ and obtained the curves by carrying out a Maxwell construction. For small values of $J$ and $U$ the curves show a smooth behavior with the particle number changing in discrete steps of 1 at a time. However, for densities close to $N_1/L=1 or 0$ and especially when $J$ is increased we find that in certain density regimes the jumps are now in steps of two. This is an indication of a pairing instability corresponding to the formation of bi-excitons. In order to determine whether these bi-excitons are stable objects in the thermodynamic limit, we need to carry out a finite size analysis of the binding energies. To distinguish different regimes we first define the binding energy for two particles pairing on each orbital chain separately as:
\begin{eqnarray}
\Delta_{\lambda=1} &=& \left(E(N_1-2,N_2)-E(N_1,N_2)\right) \nonumber \\
&-& 2\left(E(N_1-1,N_2)-E(N_1,N_2) \right)  \\
& = & E(N_1,N_2)+E(N_1-2,N_2)-2E(N_1-1,N_2) \nonumber,
\end{eqnarray}
and a similar expression for $\lambda=2$ obtained by exchanging the labels. These quantities determine whether it is energetically more costly to remove two particles, compared to twice the energy of removing one. The difference between the two indicates the binding energy, which is negative in the case of an attraction between partcles.
This idea can be generalized to the case of a particle-hole pair: the binding energy for the formation of a single exciton is given by:
\begin{eqnarray}
\Delta_{ex} &=& E(N_1,N_2)-E(N_1-1,N_2)- \\ \nonumber
 &- & E(N_1,N_2+1)+E(N_1-1,N_2+1).
\end{eqnarray}

%and for a bound pair of excitons by:
%\begin{equation}
%\Delta_{2ex}=E(N_1,N_2)+E(N_1+2,N_2-2)-2E(N_1+2,N_2-1).
%\end{equation}

\begin{figure}%[ht]
\centering
\includegraphics[width=0.48\textwidth]{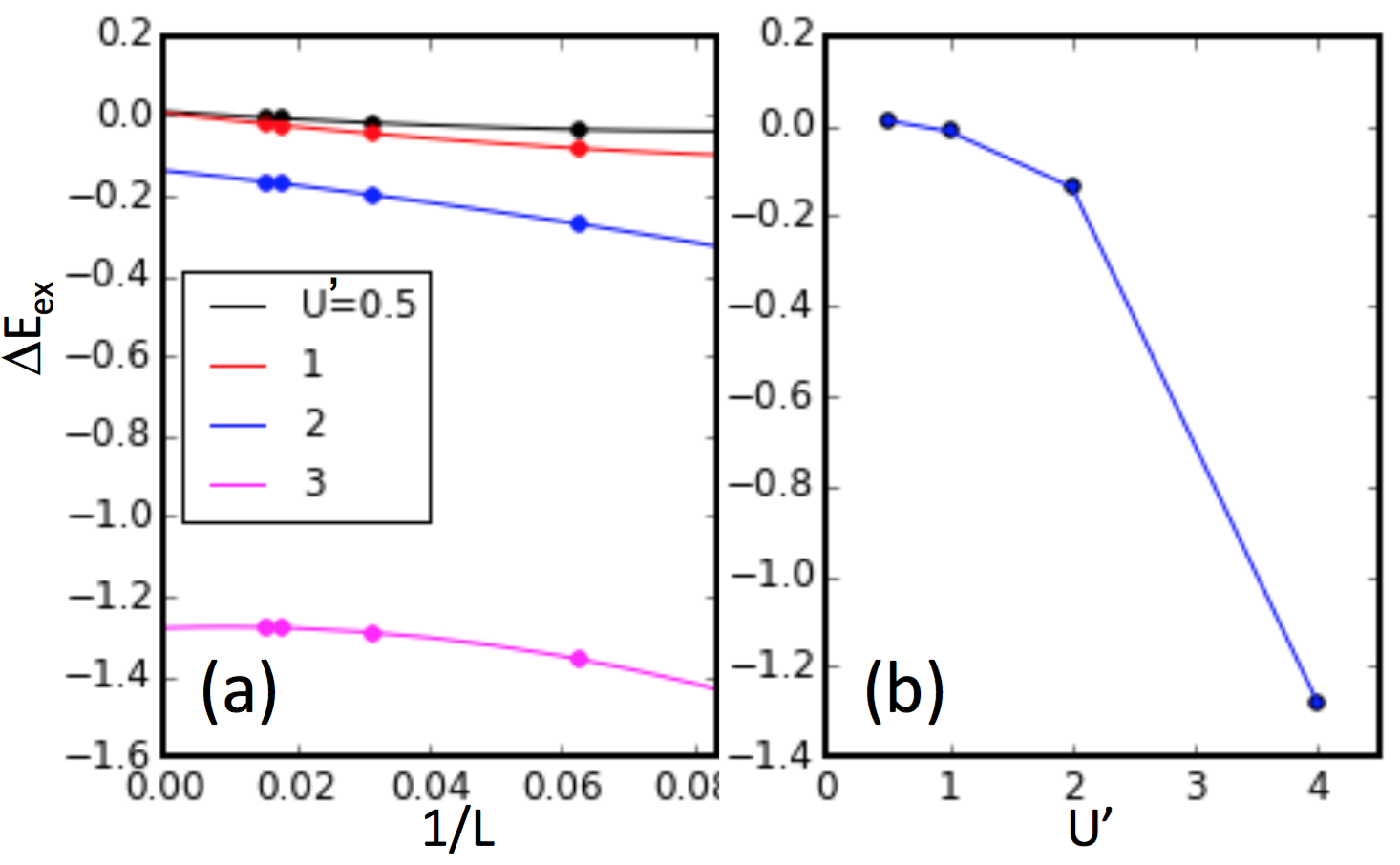}
\caption{(a) Finite size scaling of the single exciton binding energy as defined in the text for $J=1.2$ and different values of $U'$ and densities $N_1/L=3/4$ and $N_2/L=1/4$. (b) Results of the extrapolation to the thermodynamic limit as a function of $U'$. }
\label{fig:binding1}
\end{figure}

Results for several parameter regimes and system sizes are shown in Fig.\ref{fig:binding1}(a), focusing on the regime $N_1/L=0.75$. In Fig.\ref{fig:binding1}(b) we also plot the values in the thermodynamic limit, as obtained from a quadratic fit in $1/L$.
For small values of $U'$ it is difficult to tell from our results if the particle-hole excitations form or not bound states. It is also possible that the electrons in the upper band form bound singlets that propagate independently as observed in the 1D $t-J$ chain. However, this would occur for large values of $J \sim 2t$. On the other hand, increasing the value of $U'$ makes the mass of the excitons very heavy and these clump together and the system phase segregates.

\subsection{Excitonic density waves and charge order}

\begin{figure}%[ht]
\centering
\includegraphics[width=0.48\textwidth]{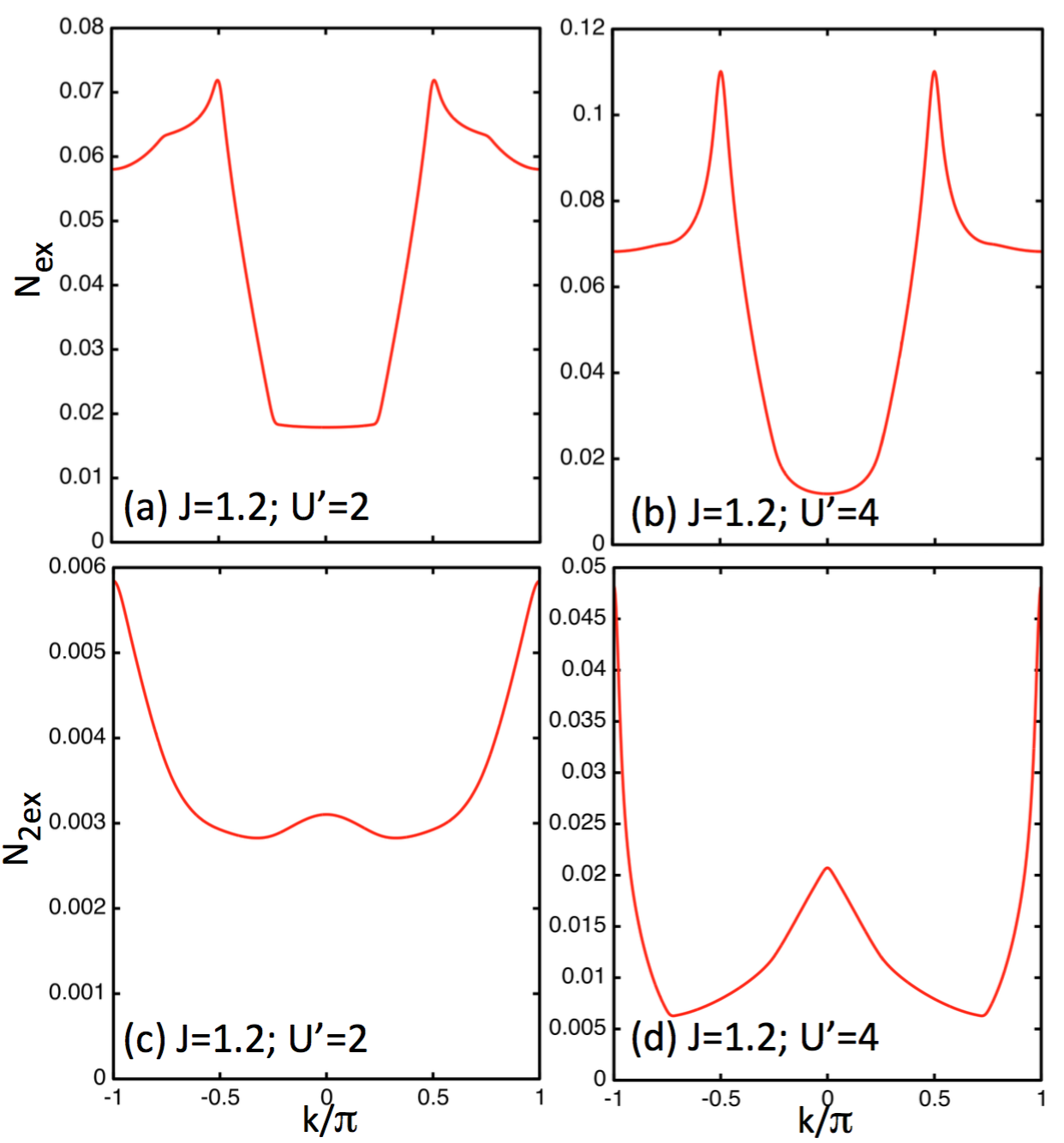}
\caption{Single exciton momentum distribution function (MDF) for $L=64$; $N_1=48$, $N_2=16$ and (a) $J=1.2, U'=2$ in the excitonic phase and (b) $J=1.2, U'=4$ in the bi-excitonic phase. Panels (c) and (d) show the bi-exciton MDF for the same parameters, respectively.}
\label{fig:mdf}
\end{figure}

\begin{figure}%[ht]
\centering
\includegraphics[width=0.48\textwidth]{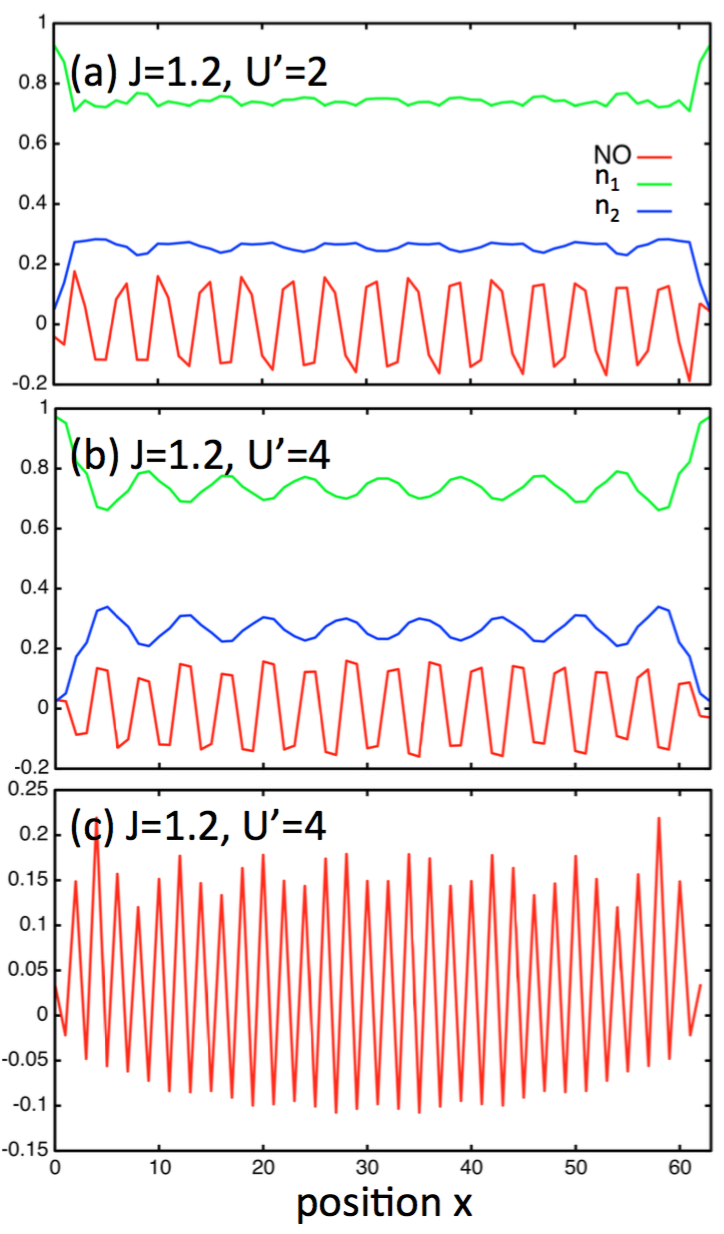}
\caption{Natural orbitals for the exciton condensate with $L=64, N_1=48; N_2=16$ in two parameter regimes: (a) $J=0.6, U'=2$ corresponding to the excitonic phase; (b) $J=1.2, U'=2$ in the bi-excitonic phase. Panel (c) shows the natural orbital for the bi-excitonic condensate. We also show the local occupation of the two orbitals, $n_1$ and $n_2$. }
\label{fig:no}
\end{figure}

In order to determine the ground state properties we study several correlation functions, paying particular attention to the cases with $J=1.2$. In Fig.\ref{fig:mdf} we plot the exciton and bi-exciton momentum distribution functions (MDF), defined as:
\begin{eqnarray}
N_{ex}(k,\sigma,\sigma') &=& \frac{1}{L}\sum_{x,y} e^{ik(x-y)}\langle b^\dagger_{x\sigma} b_{y\sigma'} \rangle \nonumber \\
N_{2ex}(k) &=& \frac{1}{L}\sum_{x,y} e^{ik(x-y)}\langle \Delta^\dagger_{x} \Delta_{y} \rangle .
\end{eqnarray}
A conventional approach in DMRG calculations with open boundary conditions consists of averaging data taken at distances that are equidistant from the center (we refer the reader to Ref.\cite{White1993b} for details). In large systems, particularly with a gap, boundary corrections are typically small. As we shall see below, in the particular cases of interest, edge effects involve very few lattice spaces (see for instance, Fig. \ref{fig:no}).
These expressions assume that the excitons are local objects that can be described in terms of bosonic operators $b^\dagger_{x\sigma}=c^\dagger_{x\sigma 2}c_{x\sigma 1}$ and that the bi-excitons can form pairs\cite{Okumura2001,Combescot2007} $\Delta^\dagger_{x}=\frac{1}{\sqrt{2}}\left(b^\dagger_{x\uparrow}b^\dagger_{x+1,\downarrow}-b^\dagger_{x\downarrow}b^\dagger_{x+1,\uparrow}\right)$. Since our model does not take into account inter-orbital hybridization nor Hund's coupling, $N_{ex}$ is always diagonal in the spin index and from now on we only consider $N_{ex}(k,\uparrow,\uparrow)$ \cite{Kaneko2014}.
It is clear beforehand that the actual excitonic wave function may actually spread over several lattice spaces, but these quantities offer a quite good description of the underlying ground state and its pairing tendencies. The excitonic MDF, for instance, shows a clear peak at $k=\pi/2$, indicating that the quasi-condensate of excitons has a finite center of mass momentum, an excitonic density wave (EDW), as anticipated. The bi-excitonic MDF shows some structure for $U'=2$ but the maximum at $k=\pi$ cannot be characterized as a peak, particularly by looking at the scale on the $y$-axis. On the other hand, the one for $U'=4$ shows a quite dramatic peak. This can be interpreted as a quasi-condensate of bi-excitons with finite center of mass momentum $Q=\pi$ formed by single exciton pairs with momentum $\pi/2$. This also gives rise to a small peak at zero momentum, but it is less defined and much broader.

These observations can be made more explicit by studying the quasi-condensate wave function by means of Penrose
and Onsager's description of the superfluid order parameter.\cite{Penrose1956}
 The natural orbitals (NO) $\psi_\alpha$ of the system will simply be the single particle
  eigenstates  -- in the bosonic sense -- of the bosonic single-particle density matrix:
\begin{eqnarray}
G_{ex}(x,y) &=& \langle b^\dagger_{x\uparrow} b_{y\uparrow} \rangle \nonumber \\
G_{2ex}(x,y) &=& \langle \Delta^\dagger_{x} \Delta_{y} \rangle .
\end{eqnarray}
%The eigenvalues $\lambda_\alpha$ represent their occupations.
 The NO with the largest eigenvalue, $\psi_0$, is the single-particle state in which quasi-condensation takes place. We generalize this concept to the case of excitons and bi-excitons and show the results in Fig.\ref{fig:no} for $U'=2$ in the excitonic phase and $U'=4$ in the bi-excitonic phase. The periodicity of the wave functions is determined by the momentum of the condensate: $Q=k_{F1}+k_{F2}=\pi/2$, and $Q=\pi$ for single excitons and bi-excitons, respectively (see Fig.\ref{fig:mdf}). 

\begin{figure}%[ht]
\centering
\includegraphics[width=0.48\textwidth]{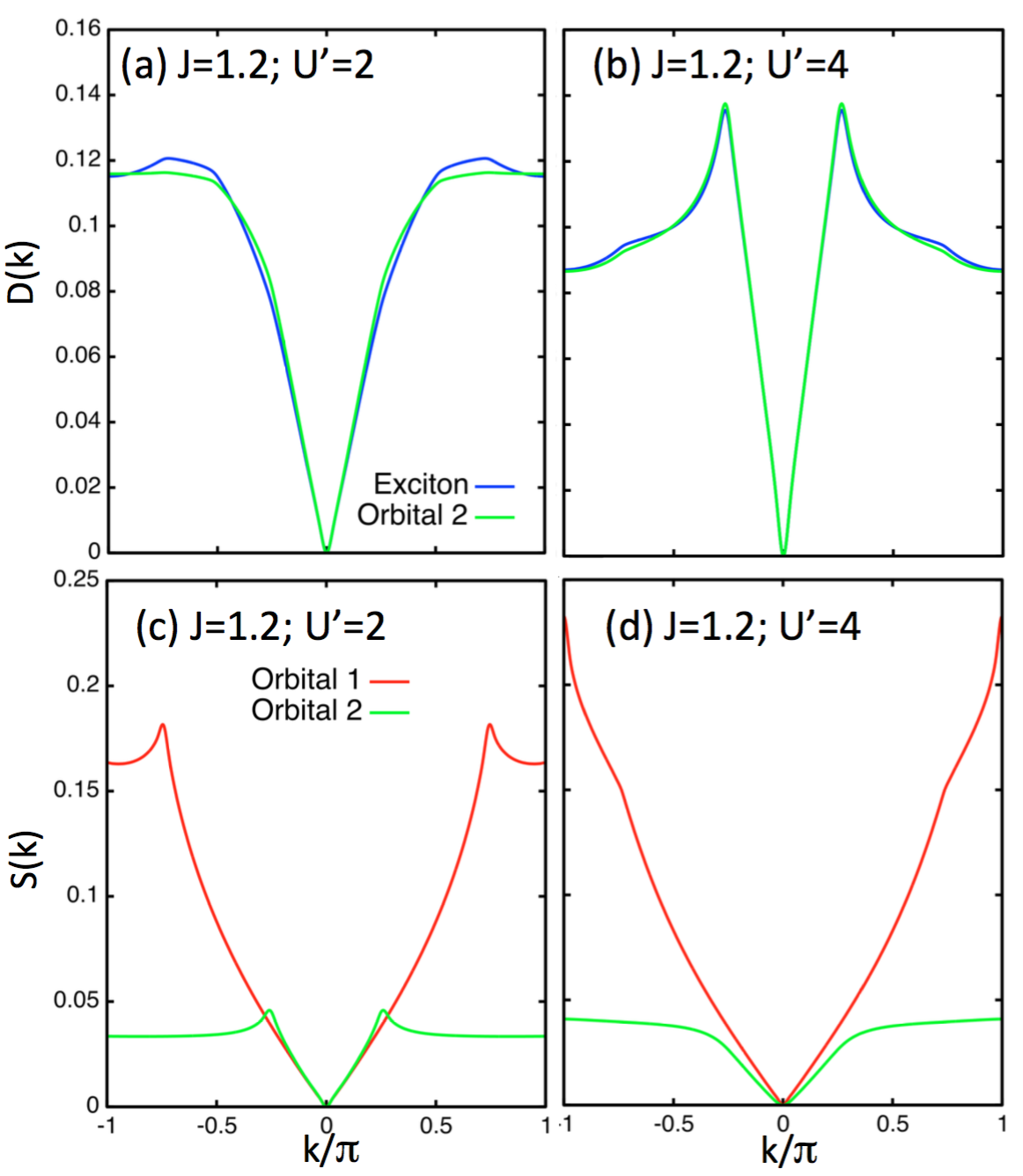}
\caption{Density structure factor for excitons and electrons in orbital $\lambda=2$ for $L=64, N_1=48; N_2=16$ and parameters (a) $J=1.2, U'=2$ in the excitonic phase and (b) $J=1.2, U'=4$ in the bi-excitonic phase. Panels (c) and (d) show the spin structure factor for the same parameters, respectively.}
\label{fig:structure}
\end{figure}

It is important to point out that a condensate with periodicity $\pi/2$ does not indicate charge order with period $\pi/2$ ({\it i.e.} ``1-1-0-0''). This would only occur at quarter filling with $N_2=N_1=L/2$. As a matter of fact, the density of excitons is not commensurate with this order. This is illustrated in Fig.\ref{fig:structure} by our results for the density-density structure factor:
\begin{equation} 
D_{\lambda}(k) = \frac{1}{L}\sum_{x,y} e^{ik(x-y)}\langle n_{x\lambda} n_{y\lambda} \rangle, 
\end{equation}
where $n_{x\lambda}=\sum_\sigma c^\dagger_{x\sigma\lambda}c_{x\sigma\lambda}$ and a similar expression for the excitonic density:
\begin{equation}
D_{ex}(k) = \frac{1}{L}\sum_{x,y} e^{ik(x-y)}\langle n_{ex,x} n_{ex,y} \rangle, 
\end{equation}
where $n_{ex,x}=b^\dagger_{x,\uparrow} b_{x,\uparrow}$ is the number operator for excitons.
The excitonic structure factor and the one for orbital $\lambda=2$ are practically indistinguishable, indicating that holes and electrons are forming tightly bound pairs.
Signatures of charge order would be identified as peaks at finite momentum. The case $U'=2$ does not show any structure and is practically featureless, as expected from a dilute condensate of hard-core bosons/excitons. On the other hand, for $U'=4$ one can clearly see the onset of charge order with momentum $2k_{F2}=\pi/4$. This resembles a state in which EDW and CDW orders coexist and are intertwined. In order to determine if this state is or not a CDW, we calculate the charge gap for adding/removing pairs of excitons. This is defined as:
\begin{equation}
\Delta_{ch}=E(N_1+2,N_2-2)+E(N_1-2,N_1+2)-2E(N_1,N_2).
\end{equation}
A finite size scaling (not shown) indicates that this quantity vanishes in the thermodynamic limit. Therefore, this state is not quite a CDW, but a condensate of bi-excitons, and the modulation observed in the charge density (Fig.\ref{fig:no}) corresponds to slowly decaying Friedel oscillations due to the open boundaries, as also observed in $t-J$ ladders\cite{White2002}.
We could have anticipated this conclusion from the density profile shown in Fig.\ref{fig:mu}: a CDW would be reflected as plateaus, which clearly are not observed.

%For illustration, we show the condensate and the local density for $L=64$, $N_1=24$ and $N_2=12$ with $U'=4, J=1.2$ in Fig.\ref{fig:low_density}.
%%%%%%%%%%%%%%%%%%%%%%%%%%%%%%%%%%%%%%%%%%%%%%%%%%

Finally, for completeness, in the same Fig.\ref{fig:structure} we show the spin structure factor:
\begin{equation}
S_{\lambda}(k) = \frac{1}{L}\sum_{x,y} e^{ik(x-y)}\langle S^z_{x\lambda} S^z_{y\lambda} \rangle.
\end{equation}
For $U'=2$, both orbitals display small peaks at $k=2k_{F\lambda}$. However, in the bi-excitonic phase the peak or orbital $\lambda=1$ has moved to $k=\pi$, while structure factor for orbital $\lambda=2$ is now completely featureless. This is expected from excitons bound into spin singlet pairs with short ranged correlations. In addition, the peak at $\pi$ indicates that the bi-excitons do not disrupt the antiferromagnetic order.

\begin{figure}%[ht]
\centering
\includegraphics[width=0.48\textwidth]{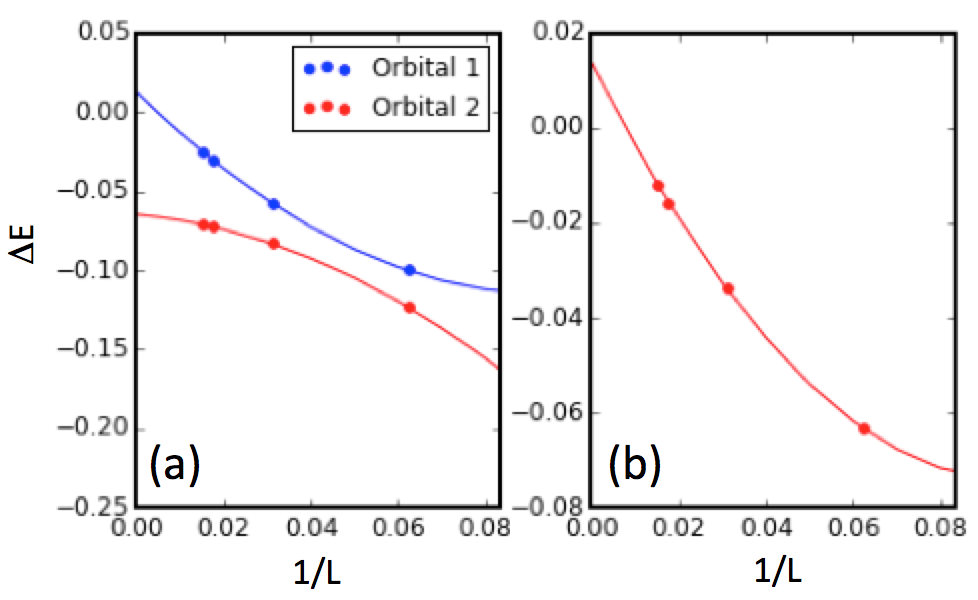}
\caption{Finite size scaling of the (a) single particle binding energy for each orbital chain and (b) exciton binding energy. Results are for $J=1.2$ and $U'=1$ and density $N_1/L=0.75$.}
\label{fig:orbital_selective}
\end{figure}

\subsection{Orbital selective pairing}

An additional feature of our model is that it naturally realizes a phase in which one of the orbitals behaves as a Luttinger liquid, while the second one undergoes a pairing instability. 
For small $U'$, the orbitals are practically decoupled and our model behaves as two independent $t-J$ chains. 
At relatively large values of $J \sim 2t$ and low densities the $t-J$ chain presents a singlet-superconducting phase with a spin gap\cite{Moreno2011}. Therefore, one can tune the parameter $\Delta$ such that the occupation of each orbital falls into a different phase. This occurs for instance for $U'=0.5, J=2.4, N_1/L=0.75, N_2/L=0.25$. In Fig.\ref{fig:orbital_selective} we show that the binding energy for the low-density chain is finite, while it remains positive for the high-density one. In addition, the binding energy for exciton formation is also positive. This description offers a simple and natural scenario for the realization of this type of orbital selective paired states.

\subsection{Away from quarter filling}

\begin{figure}%[ht]
\centering
\includegraphics[width=0.48\textwidth]{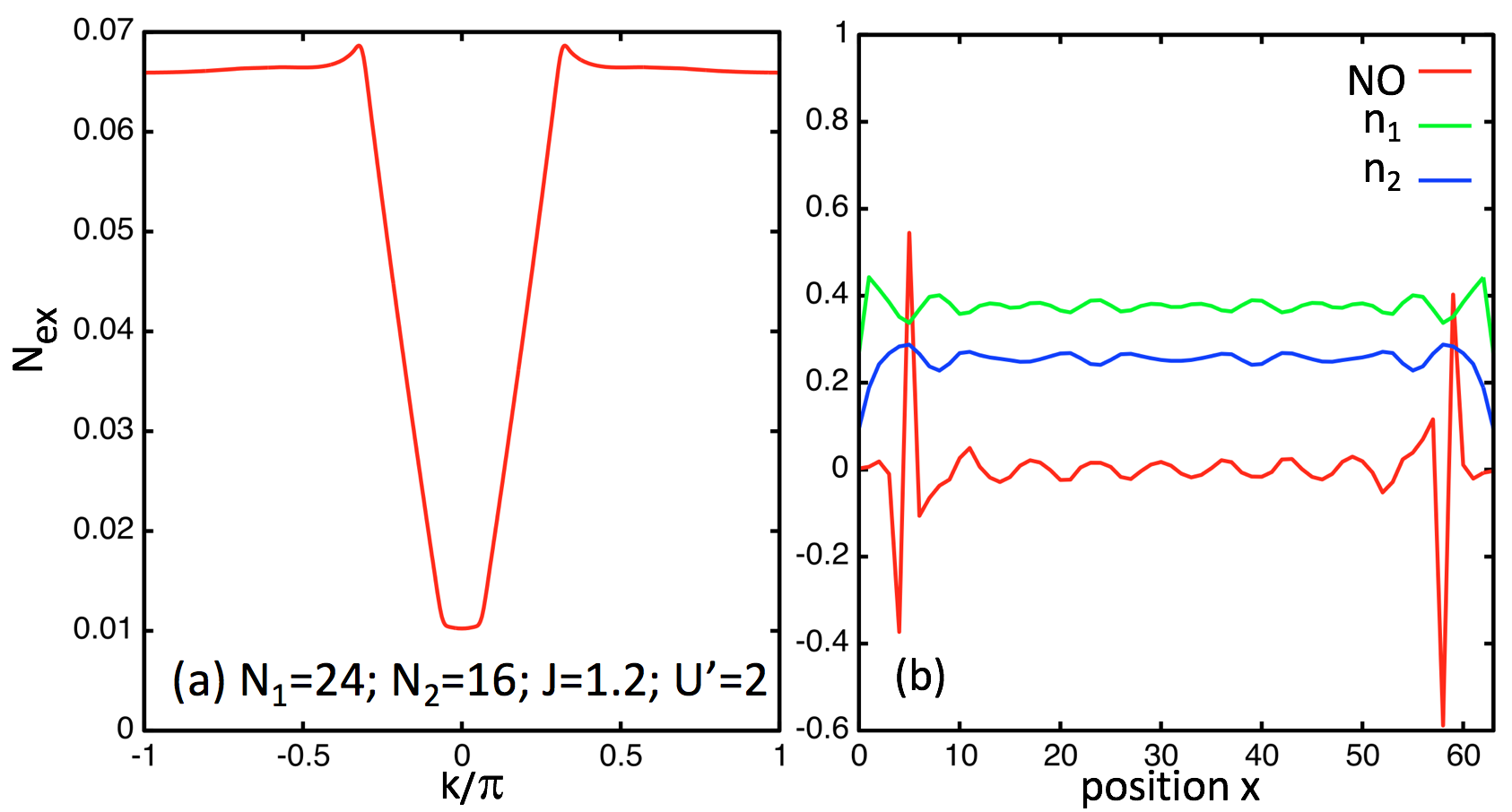}
\caption{(a) Excitonic momentum distribution function for $L=64$, $N_1=24$, $N_2=16$, $J=1.2$ and $U'=2$. (b) Natural orbital for the exciton condensate and the local occupation of the two orbitals, $n_1$ and $n_2$, as in Fig.\ref{fig:no}. The edge effects are due to the open boundary conditions. }
\label{fig:away}
\end{figure}

The excitonic physics discussed for the quarter filling case extends to other filling fractions as well. Without attempting to determine a phase diagram, we just show some typical results that we obtained for small densities in Fig.\ref{fig:away}. As shown in panel (a), the exciton MDF is peaked at a finite value of $Q=k_{F1}+k_{F2}$, which is reflected in the behavior of the natural orbitals, displayed in panel (b). The charge structure factor (not shown) indicates a state with no charge order. In our exploration of parameter space we have not found bi-excitonic physics, but this may appear at larger values of $J$ than the ones we considered. An extended study is currently underway and will be presented elsewhere. 

\section{Conclusions}\label{conclusions}

\begin{figure}%[ht]
\centering
\includegraphics[width=0.48\textwidth]{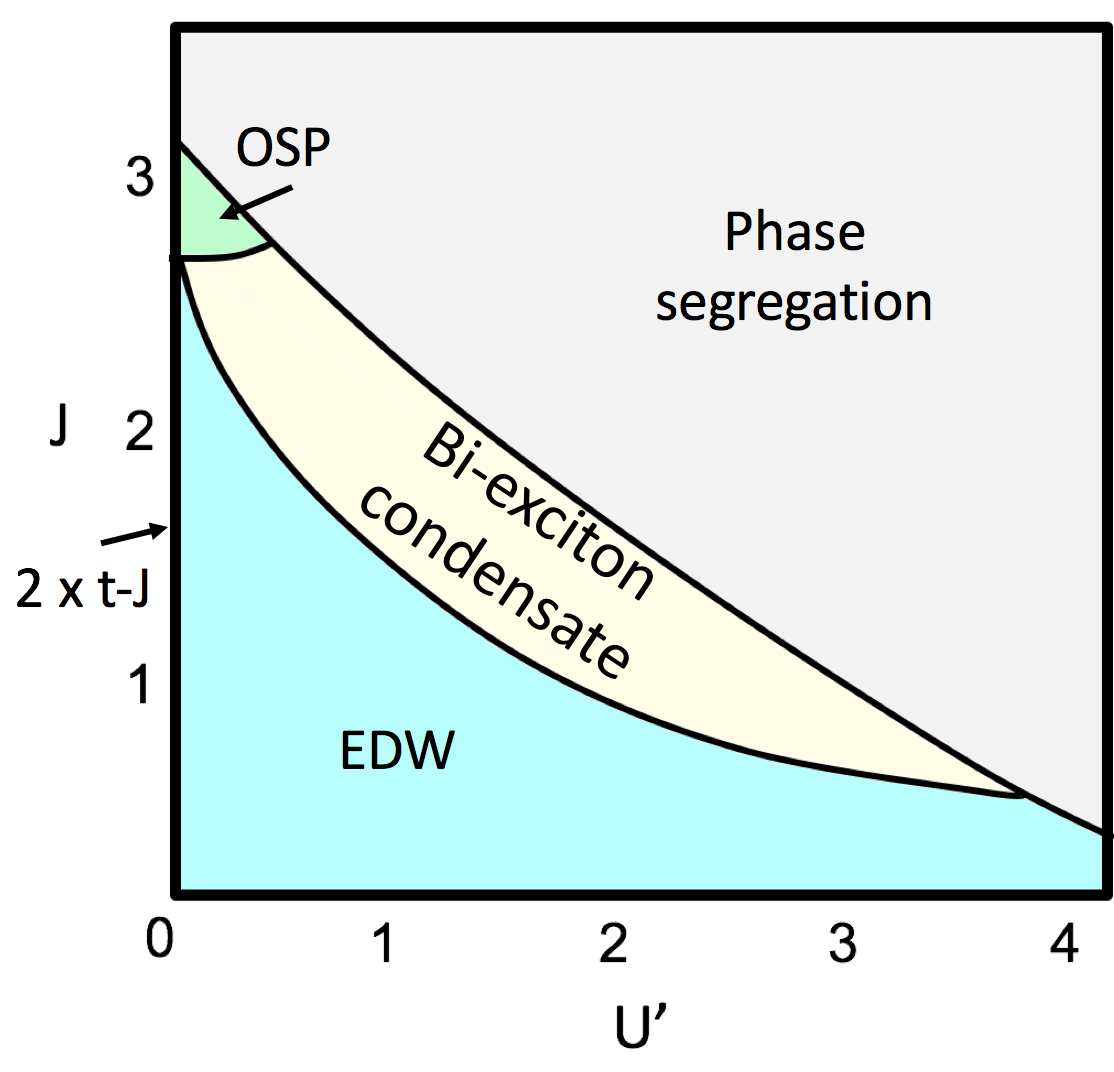}
\caption{Schematic phase diagram of the two orbital model as a function of $U'$ and $J$ for fixed densities $N_1/L=3/4$, $N_2/L=1/4$. Along the $U'=0$ line the system consists of 2 copies of a $t-J$ chain at different densities. 
Finite values of $U'$ induce the formation of an exciton density wave (EDW), and increasing $J$ drives an instability toward pairing of excitons (bi-exciton condensate).
At small values of $U'$ and large values of $J$ we find the orbital selective paired phase (OSP). 
}
\label{fig:pd}
\end{figure}

We have presented a detailed study of the exciton and bi-exciton formation in a one-dimensional two-orbital $t-J$ model. The stability of the excitons is determined by the strength of the inter-orbital Coulomb interaction $U'$, while the formation of bi-excitons is controlled by the antiferromagnetic exchange $J$. A schematic phase diagram for densities $N_1/L=3/4; N_2/L=1/4$ is shown in Fig.~\ref{fig:pd}. For weak $U'$ the system behaves as two independent decoupled chains. It is possible that the system is inherently unstable to exciton formation and this occurs for any finite $U'$. This would correspond to an exciton binding energy that grows exponentially with $U'$, something difficult to resolve even with a careful finite size analysis. Nevertheless, as $U'$ is increased, we find an instability toward exciton formation such that they form a quasi-condensate with finite center of mass momentum, corresponding to an excitonic density-wave. This can be understood through our analysis of the three-body problem of an electron-hole pair and a spinon: at quarter-filling the system behaves basically as a single doped $t-J$ chain where the excitons act as holes hopping with both nearest, and next-nearest hoppings. These holes are heavier and can condense, since in reality they are electron-hole bound states. 

In general, the period of the EDW will be determined by the excitonic fraction $N_2/L$ (or $\Delta$). It is important to point out that this state does not correspond to a CDW (or excitonic CDW), since there is no charge order. 
Notice that the condensate wavefunction, or natural orbital, alternates signs as ($++--$) like a square wave that has no nodes. Therefore, the probability density, which is the square of it, also has no nodes and, moreover, it is not commensurate with the density, hindering the possibility of an FFLO-like  phase.  

As the interactions $U'$ are increased, the excitons become heavier and more localized enabling the exchange interaction to bind them into bound pairs. At the same time we observe signatures of an instability toward a charge density-wave of bi-excitons, reminiscent of the idea of an excitonic crystal\cite{Ogawa2007a,Ivanov1993}. However, bi-excitons are not localized and the period of the CDW is different than the period of the condensate. Since the charge gap vanishes, we conclude that this is not CDW but a condensate of bi-excitons.
%The coexistence of bi-excitons with charge ordering is reminescent of recent proposal to describe the pseudogap phase in cuprate superconductors. 
For large values of the parameters, the system phase separates. This can occur in two different ways: (i) For large $U'$ the system splits into electron-rich and hole-rich domains; within each domain, each orbital forms a Mott insulating Heisenberg chain. (ii) For small $U'$ and large $J$ we find the physics of two $t-J$ chains that phase segregate independently, as encountered in the phase diagram of the single-orbital problem \cite{Moreno2011}. Before this occurs, though, we find a regime around $J \sim 2t$ in which one orbital is metallic, while the other one is a spin gapped superconductor, an actual orbital selective paired state.

The observed excitonic density wave can be directly related to pair density waves \cite{Zachar2001,Berg2010,Fradkin2015} in a very simple way: a particle-hole transformation in the high-energy orbital $\lambda=2$ leads to a one-to-one correspondence between excitons (neutral particle hole pairs) and Cooper pairs (with charge $2e$), with the excitonic condensate translating into a pair density wave. The parent Hamiltonian of this state would have negative $U'$ and would pair electrons with momentum $k_{F1}$ and $-k_{F2}$, identical to what takes place in the FFLO phase of the negative $U$ Hubbard chain\cite{Feiguin2007c,Feiguin2009b,Heidrich-Meisner2010,Dalmonte2012}. Moreover, the bi-excitonic regime would correspond to a PDW of composite objects of charge $4e$ similar to predictions for stripe superconductors in Ref.~\onlinecite{Berg2009}. In the language of hard-core bosons this would correspond to a condensate of bosonic pairs with finite center of mass momentum.

This behavior occurs also at densities below quarter filling. The center of mass momentum for the EDW is given by $Q=k_{F1}+k_{F2}$ and can acquire a long wavelength when this difference is small. 
 
The model displays rich physics with a number of phases that resemble the phenomenology of both cuprates and iron pnictides, encouraging us to believe that there is much to learn from multi-orbital model Hamiltonians that can guide our intuition toward a comprehensive picture of these materials. Needless to say, one can expect yet richer physics once the Hund interaction is taken into account\cite{Nonne2010}. 

\acknowledgments
We are greateful to R.~ Markiewicz and E. Fradkin for illuminating discussions.
The authors acknowledge the U.S. Department of Energy, Office of Basic Energy Sciences for support under grant No. DE-SC0014407. 

%\bibliography{reviews,dmrg,career,supersolid,mypapers,pdw,si-gs,bound,excitons,excitons2,excitons3,laser,polymers,spin-orbital-separation,orbital-selective,xrays}

\end{document}